\documentclass[reprint,amsmath,amssymb,aps,prx,floatfix]{revtex4-2}
\usepackage{graphicx}
\usepackage{dcolumn}
\usepackage{bm}
\usepackage{hyperref}
\usepackage{balance}

\begin{document}

\preprint{APS/123-QED}

\title{Overcoming Fourier Locking in Quantum Data Re-uploading Classifiers\texorpdfstring{\\}{ }via Spectral Homotopy}
\author{Spencer Topel}
\email{spencer@mothquantum.com}
\affiliation{Moth, Brooklyn, New York}

\date{\today}

\begin{abstract}
Data re-uploading parameterized quantum circuits (DRU-PQCs) are universal function approximators, yet their expressivity produces oscillatory, non-convex loss landscapes that resist gradient-based optimization. We show that the primary optimization bottleneck in DRU-PQCs is not insufficient capacity but a structural failure mode we term \textit{Fourier locking} (FL): because encoding weights and entangling layers are nonlinearly coupled, random initialization on high-frequency targets collapses the encoding parameters into spurious local minima. Two Fisher diagnostics characterize FL. The input-space quantum Fisher information $F_x$ measures the effective frequency content of the encoded state; the Fisher discriminant ratio of the measured features measures their alignment with the class labels. In two independent 50-seed experiments, the locking is literal: trapped circuits hold $F_x$ frozen for the entire run, while escaping circuits migrate their frequency content (direct training: $r_{pb} = -0.48$; curriculum: $d = 1.34$; both $p < 0.001$). The replicated signature is this spectral mobility, not any endpoint value of $F_x$, and trapped circuits retain a fully non-degenerate parameter-space QFIM ($r_{pb} \approx 0$): the failure is spectral misalignment of a responsive state, not a loss of geometric sensitivity. A \textit{frequency-staged homotopy} protocol that paces the target frequency ($f\colon 1.0 \to 3.0$) convexifies the early loss landscape; escaping circuits raise $F_x$ in step with the curriculum, and the escape rate triples (18\% vs.\ 6\%). Fourier locking is a frequency-alignment problem, and its remedy is frequency pacing.
\end{abstract}
 
\maketitle

\section{Introduction}\label{sec:intro}
In current literature, parameterized quantum circuits utilizing data re-uploading (or DRU-PQCs) are often assumed to be advantageous due to their universal expressivity. However, as noted in recent theoretical work, high expressivity degrades trainability, generating highly oscillatory, non-convex loss landscapes \cite{holmes2022connecting}.

In this paper we demonstrate that the primary optimization bottleneck in DRU-PQCs is not a deficit in expressive capacity. Instead, we argue it is a spectral failure mode we term Fourier locking (FL). Because encoding weights (frequency selectors) and entangling layers (signal routers) are non-linearly coupled, random initialization on high-frequency targets causes the encoding parameters to collapse into isolated, spurious local minima.

A contribution of this paper is demonstrating that standard gradient magnitudes fail to diagnose FL traps, and that the natural quantum-geometric suspects fail as well. We measure the parameter-space Quantum Fisher Information Matrix (QFIM) throughout training and find it does not collapse in trapped circuits: the locked model remains fully sensitive to its parameters. Instead, two quantities carry the diagnostic signal. The Fisher discriminant ratio (FDR) of the measured features collapses when a circuit locks, identifying the trap as a failure of label alignment at the readout. And the input-space quantum Fisher information $F_x$---the Fubini--Study susceptibility of the state to the encoded data---identifies its dynamical signature: the effective frequency content of a trapped circuit is frozen at a misaligned value from initialization onward, while escaping circuits migrate theirs over training.
 
To test this mechanism we deploy frequency-staged homotopy, sequentially pacing the target frequencies ($f=1.0 \rightarrow 3.0$) to smooth the initial loss landscape. Rather than presenting this pacing as a definitive training solution, we use it as a probe of the frequency-alignment picture. Tracking $F_x$ across the curriculum reveals a distinct latent scaffold mechanism: circuits that escape grow their effective frequency content in step with the curriculum, while trapped circuits sit at a static, misaligned frequency from initialization onward. This supports the claim that FL can be bypassed if the encoding is transported into the target frequency band before the high-frequency attractors materialize.

\section{Background}\label{sec:background}
\subsection{Quantum Data Re-uploading and Universal Expressivity}
A known challenge in Quantum Machine Learning (QML) is efficient encoding of classical data into quantum states. Early variational quantum circuit models relied on single-instance encoding that limited the complexity of the functions they could approximate. To overcome this, P\'{e}rez-Salinas et al. \cite{perez2020data} introduced a quantum data re-uploading architecture. By sequentially injecting the same classical data into the rotation angles of single-qubit gates across multiple layers, their architecture mimics the deep, non-linear processing of classical neural networks. This repetitive encoding allows even a single-qubit circuit to act as a universal classifier, capable of drawing complex, non-linear decision boundaries.

\subsection{The Fourier Representation of Quantum Models}
To understand the mathematical source of this expressivity, Schuld et al. \cite{schuld2021effect} demonstrated that variational quantum machine learning models effectively compute Fourier series. In this framework, the data encoding strategy dictates the available frequencies, while the parameterized entangling layers dictate the coefficients of those frequencies.

Crucially, every time data is re-uploaded into the circuit it expands the frequency spectrum available to the model. The so-called expressive power of a Data Re-uploading Parameterized Quantum Circuit (DRU-PQC) is directly proportional to its depth: deeper circuits with more re-uploads can characterize oscillatory and high-frequency data features.

\subsection{The Expressivity Trap and Barren Plateaus}
While high expressivity is theoretically desirable, recent literature has established that it actively sabotages optimization. In classical networks, over-parameterization often smooths the loss landscape. In quantum circuits, however, high expressivity creates oscillatory, non-convex landscapes and induces barren plateaus \cite{holmes2022connecting}.

Recent theoretical and empirical analyses have shown that as encoding depth increases, trainability sharply degrades, introducing barren plateaus and reducing predictive performance to near-random guessing without a corresponding gain in expressibility \cite{alamo2025cheat, wang2025predictive}. Specifically, when tasked with high-dimensional or high-frequency data, deep re-uploading circuits are susceptible to barren plateaus and severe predictive degeneration. The landscape becomes computationally intractable to navigate in the worst case, causing models to collapse into a degenerate ``mean-prediction'' state where outputs become invariant to the input data, effectively reducing accuracy to random chance.

\subsection{FL and Trapped Gradient Descent}
Quantum Machine Learning (QML) literature often attempts to overcome optimization bottlenecks by dynamically expanding architectural capacity. However, theoretical bounds establish that training Variational Quantum Algorithms is, in the worst case, an NP-hard problem \cite{bittel2021training}. Importantly, this bound applies to the worst-case optimization problem over all possible landscapes. Moving beyond these formal computational limits, our empirical analysis identifies FL as the specific, typical-case geometric mechanism responsible for these failures in Data Re-uploading PQCs. 

In these expressive architectures, encoding weights (which dictate the input frequency) and entangling gates (which route the quantum information) are non-linearly coupled. When randomly initialized on a high-frequency target, the optimizer is thrust into an oscillatory, non-convex loss landscape. Because standard gradient descent is short-sighted and follows local slopes, FL occurs when these gradients trap the encoding weights in a spurious, high-frequency minimum early in training. Once the encoding parameters commit to a misaligned attractor, the entangling layers cannot route a signal that effectively does not exist.

This spectral failure produces a diagnostic signature, but not the one geometric intuition suggests. The analytical gradients vanish ($\nabla_\theta L \approx 0$), yet the parameter-space QFIM of the trapped circuit does not collapse: the state remains fully responsive to its parameters, and its input-space QFI shows it oscillating strongly with the data. What collapses is the Fisher discriminant ratio of the measured features---the class alignment of the readout. Unlike standard barren plateaus---where landscapes flatten through the chaotic expressivity of Haar-random states---the Fourier lock produces a flat loss because the state's oscillations occur at frequencies the measurement cannot map onto the class labels. The optimizer is paralyzed not by a lack of representational capacity, nor by a severing of the parameter geometry, but by a structural misalignment between the frequencies the encoding has selected and those the target boundary requires, necessitating an intervention that modifies the landscape before gradient descent can succeed.

\section{Theoretical Framework}\label{sec:framework}

\subsection{The DRU-PQC as a Fourier Series}
We must first formalize the mathematical structure of the Data Re-uploading Parameterized Quantum Circuit (DRU-PQC) and the diagnostic metrics used to evaluate its optimization landscape. 

Following the framework established by Schuld et al. \cite{schuld2021effect}, a DRU-PQC does not merely process data; it constructs a Fourier series. For a 4D input $x$, the action of a single data-encoding gate with a trainable encoding weight $\omega$ can be expressed as $R(x, \omega) = e^{-i \omega x P}$, where $P$ is a Pauli generator.

When this input is sequentially re-uploaded across $L$ layers interspersed with trainable entangling blocks $W(\theta)$, the final output of the quantum model—measured as the expectation value of an observable $O$—takes the form of a truncated Fourier series:
\begin{equation}
f(x, \omega, \theta) = \sum_{k \in \Omega} c_k(\theta) e^{i k \omega x} 
\label{eq:fourier}
\end{equation}
Here, the accessible frequency spectrum $\Omega$ is determined by the encoding weights $\omega$ and the number of re-uploading layers $L$. The complex coefficients $c_k(\theta)$ are determined by the entangling parameters $\theta$.

\subsection{The Joint Parameter Bottleneck}
Equation (\ref{eq:fourier}) reveals the fundamental optimization bottleneck unique to DRU-PQCs: the non-linear coupling between $\omega$ and $\theta$.

In a classical neural network, weights independently transform the input. In the DRU-PQC, $\omega$ dictates which frequencies are available, while $\theta$ dictates how those frequencies are routed to form a decision boundary. If the encoding weights $\omega$ are initialized such that the target frequencies are heavily suppressed or excluded from the spectrum $\Omega$, the entangling parameters $\theta$ are effectively blind. They cannot optimize a signal that the encoding layers have failed to capture. This structural dependency is the mathematical root of FL.

\subsection{Two Fisher Diagnostics: Readout Alignment and Input-Space Geometry}
\label{sec:diagnostics}
 
Because of the joint parameter bottleneck, standard gradient magnitudes are a fundamentally ambiguous metric for diagnosing landscape traps in DRU-PQCs. While gradients do vanish ($\nabla_\theta L \approx 0$) when the model is trapped in a degenerate `mean-prediction' state, this vanishing does not reveal the cause of the trap. We therefore track three Fisher quantities with distinct geometric content.
 
\textbf{Parameter-space QFIM.} The Quantum Fisher Information Matrix over the trainable parameters,
\begin{equation}
F_{ij} = 4\,\text{Re}\left( \langle \partial_i \psi | \partial_j \psi \rangle - \langle \partial_i \psi | \psi \rangle \langle \psi | \partial_j \psi \rangle \right),
\label{eq:qfi}
\end{equation}
measures the distinguishability of the state under parameter variation, as derived from the Fubini--Study metric (Appendix~\ref{app:qfi}). In the overparametrization theory of Larocca et al.~\cite{larocca2023theory}, the achievable rank of this matrix, set by the dimension of the dynamical Lie algebra, defines a static capacity ceiling. A natural hypothesis is that FL manifests as a dynamic collapse of this matrix during training. We test this hypothesis directly and it fails: as reported in Sec.~\ref{sec:results}, the normalized trace $\frac{1}{N}\text{Tr}(F)$ of trapped circuits remains of order unity throughout training and carries no diagnostic signal for the trap ($r_{pb} \approx 0$). Locked circuits are not parameter-insensitive.
 
\textbf{Input-space QFI.} The same Fubini--Study construction applied to the encoded data rather than the parameters gives
\begin{equation}
F_x = 4 \sum_{i} \left( \langle \partial_{x_i} \psi | \partial_{x_i} \psi \rangle - |\langle \psi | \partial_{x_i} \psi \rangle|^2 \right),
\label{eq:xqfi}
\end{equation}
where $\partial_{x_i}$ acts through the encoding gates. $F_x$ measures the sensitivity of the quantum state to the input, and it is controlled directly by the encoding weights: for a single encoding gate $R_X(\omega_0 x + \omega_1)$ acting on $|0\rangle$, one finds exactly $F_x = \omega_0^2$. The encoding weights are frequency selectors in Schuld et al.'s Fourier picture~\cite{schuld2021effect}, and $F_x$ is the geometric readout of the frequency they have selected. We refer to $F_x$ as the effective frequency content of the encoded state.
 
\textbf{Fisher discriminant ratio.} Finally, we track the classical Fisher discriminant ratio (FDR) of the measured features---the ratio of between-class to within-class variance of the Pauli-$Z$ expectation values over a validation set. The FDR measures whether the state's input dependence is aligned with the class labels at the readout. It is a classical statistic of the measurement outcomes, not a property of the state geometry, and the distinction matters: as we show, a locked circuit retains large $F_x$ (the state responds strongly to the input) while its FDR collapses toward zero (the response is uninformative about the labels).
 
This decomposition isolates the mechanism of the Fourier lock. When the encoding weights $\omega$ commit to a spurious high-frequency configuration, the state continues to oscillate with the input---but at frequencies mismatched to the target decision boundary. The measurement projects these oscillations onto class-uninformative directions, the FDR collapses, and the loss gradients, which are proportional to the class signal in the measured features, die with it. The failure is spectral misalignment of a fully responsive state; no degradation of the parameter geometry is required to paralyze the optimizer.

\subsection{The Mathematical Basis for Spectral Homotopy}
Standard gradient descent updates parameters via $\theta_{t+1} = \theta_t - \eta \nabla L(\theta_t)$. When applied to the high-frequency target landscape ($f=3.0$), the extreme expressivity of the Fourier series generates a highly non-convex, oscillatory loss function $L$, causing local gradient updates to immediately collapse into isolated, sub-optimal basins (Fourier locks).

We hypothesize that this can be mitigated using a continuation method in the classical sense~\cite{bengio2009curriculum}. By defining a sequence of loss functions $L_f$ parameterized by the target frequency $f$, spectral homotopy transforms the optimization objective over the training duration:
\begin{equation}
L_{f=1.0} \rightarrow L_{f=2.0} \rightarrow L_{f=3.0}
\label{eq:homotopy}
\end{equation}
The low-frequency landscape $L_{f=1.0}$ acts as a convexified surrogate. It holds the encoding weights in a trainable, low-frequency configuration---a latent scaffold---and transports the joint parameters $(\omega, \theta)$ toward the target frequency band as the curriculum advances, before the rugged terrain of $L_{f=3.0}$ is introduced.

\section{Experimental Design}\label{sec:expDesign}
In the expressive quantum data re-uploading classifiers described in Section~\ref{sec:background}, standard gradient descent is inherently vulnerable to FL: a spectral trap in which the encoding weights collapse into isolated, high-frequency local minima misaligned with the target boundary. We aim to show that this locking mechanism can be bypassed via spectral homotopy. By staging the target frequencies ($f=1.0 \rightarrow 3.0$), we begin optimization on a smoother, lower-frequency landscape, scaffolding the latent parameters into the global basin before the rugged, high-frequency attractors materialize.

To demonstrate this mechanism and map the bimodal probability space of the optimizer, the experimental design is structured to isolate the joint parameter bottleneck and test the spectral effects of frequency ordering.

\subsection{Architecture: The DRU-PQC and the Joint Bottleneck}
The core architecture is a fixed-layer, 4-qubit Data Re-uploading Parameterized Quantum Circuit (DRU-PQC). Unlike standard models that encode data once, this circuit sequentially re-injects classical inputs into the rotation angles of single-qubit gates, interspersed with parameterized entangling layers.

This design inherently couples two distinct parameter classes:
\begin{itemize}
    \item \textbf{Encoding Weights ($\omega$):} Which scale the input data, effectively acting as frequency selectors that determine the model's Fourier spectrum.
    \item \textbf{Entangling Parameters ($\theta$):} Which route the resulting quantum states through the high-dimensional Hilbert space to form decision boundaries.
\end{itemize}
The 4-qubit fixed-layer architecture possesses sufficient representational capacity to solve the target classification tasks, as the escaped runs demonstrate directly by reaching test accuracies of $0.94$--$0.96$ (Sec.~\ref{sec:results}). Any optimization failure is therefore a failure of gradient descent to navigate the parameters, not a deficit in architectural degrees of freedom.

\subsubsection{Implementation and Reproducibility}
To ensure full reproducibility of the optimization dynamics, we define the specific hyperparameter configuration of the evaluated models. The core architecture is a 4-qubit DRU-PQC comprising $L = 2$ sequential re-uploading layers. The quantum circuit contains 40 trainable weights, structured into encoding scale/bias parameters and entangling rotation angles utilizing the standard PennyLane Strongly Entangling Layers template. The quantum measurement outputs are fed into a classical linear layer (10 parameters), yielding a total hybrid model capacity of $N = 50$ trainable parameters. All models were optimized using the Adam optimizer with a learning rate of $\eta = 0.01$. The loss landscape was generated using a standard binary cross-entropy objective function across the 120-epoch training duration. A comprehensive formalization of the unitary operations and exact hyperparameter values is provided in Appendix~\ref{app:arch}.

\subsection{Diagnostic Metrics}
To diagnose the optimization dynamics without relying on potentially misleading output loss or gradient magnitudes, we track the three Fisher quantities of Sec.~\ref{sec:diagnostics} at fixed milestones throughout training.

Specifically, we track on a fixed probe batch (the first eight validation inputs) the parameter-space QFIM scalar $\frac{1}{N} \text{Tr}(F)$ of Eq.~(\ref{eq:qfi}), where $N$ is the total number of parameters; the input-space QFI $F_x$ of Eq.~(\ref{eq:xqfi}); and, over the full validation set, the Fisher discriminant ratio (FDR) of the measured features. Their distinct geometric content is developed in Sec.~\ref{sec:diagnostics}: the first measures the parameter sensitivity of the state, the second its effective frequency content, and the third the label alignment of the readout.

In expressive QML landscapes, the death of analytical gradients ($\nabla_\theta L \approx 0$) is a symptom, not a diagnosis. Tracking all three quantities disambiguates its cause: a collapse of the parameter-space scalar would indicate geometric severing, a static $F_x$ indicates frozen spectral content, and a collapsing FDR indicates a class-degenerate readout. These trajectories are tracked continuously, particularly across phase transition boundaries.

\subsection{Control Intervention: Routing Ablation}
To establish that Fourier locking (FL) is driven by the initialization of encoding weights rather than entangling logic, the first experimental phase ablates the routing mechanism. If the failure mode is merely routing inefficiency, architectural interventions targeting the entangling layers should theoretically rescue performance. To test this, we evaluate the 4-qubit DRU-PQC under three distinct initialization and routing conditions:

\begin{description}
    \item [Standard Condition (STD)] Random initialization of both the encoding weights ($\omega$) and the entangling parameters ($\theta$). 
    \item [Identity Intervention (ID)] Entangling layers are initialized to approximate Identity operations, encouraging an uncorrupted signal passthrough at $t=0$. 
    \item [Coupled Warmup Intervention (COUP)] Entangling layers undergo an independent warmup optimization phase prior to full joint training, attempting to pre-align the routing logic. 
\end{description}

Evaluating the Fisher diagnostics and escape rates across these three conditions isolates the joint parameter bottleneck, demonstrating whether the entangling layers can overcome a randomly initialized encoding state.

\subsection{Spectral Homotopy: Curriculum Ablation}
To test the core hypothesis—that explicit spectral scaffolding is required to prevent Fourier locking (FL)—the training regimen is divided into three 40-epoch phases (120 epochs total).

\textbf{Dataset Generation}\\
The target datasets are synthetically generated to isolate specific frequency learning across multiple dimensions. For a given target frequency $f$, 4D input data points $X$ are uniformly sampled from the interval $[-\pi, \pi]^4$ for a total of 4,000 samples per generated dataset (3,200 training, 800 validation). In the curriculum experiments, the training set is regenerated at the start of each phase at the active phase frequency, while the 800-sample validation set is fixed at the target frequency $f = 3.0$ throughout. The binary class labels $y \in \{0, 1\}$ are assigned using a deterministic 4D thresholding function based on the product of the target Fourier components:
\begin{equation}
y = \begin{cases} 1 & \text{if } \prod_{i=0}^3 \sin(f \cdot X_i) > 0 \\
0 & \text{otherwise} \end{cases}
\end{equation}
This formulation ensures that the geometric complexity of the decision boundary scales with the defined frequency $f$. This allows for the precise isolation of the $f=1.0$, $f=2.0$, and $f=3.0$ target topologies, with $f=3.0$ designated as the primary high-frequency target. All models were trained using a batch size of 256.

\textbf{Evaluation Framework}\\
To account for the initialization lottery, all conditions are evaluated across 50 independent random initializations (seeds). The diagnostic-instrumented experiment (Sec.~\ref{sec:bottleneck}, Fig.~\ref{fig:qfi_diagnostic}), the routing ablation, and the curriculum experiments were run with three independent seed sets; consequently, the STD baseline escape rate differs among them (8\%, 12\%, and 6\%, respectively), consistent with binomial fluctuation of a rare event across independent draws. Since performance in this landscape is fundamentally non-convex and bimodal, it is analyzed by evaluating the escape rate (defined as a final test accuracy $> 0.65$) rather than a simple arithmetic mean. The escape rate is robust to the precise threshold: escape counts are identical at thresholds of $0.60$, $0.65$, and $0.70$ in each of the four curriculum conditions, whose accuracy distributions contain no seeds between the chance cluster and $\sim$$0.70$. The diagnostic and routing seed sets contain four seeds in the $0.60$--$0.70$ interval (out of 250 runs), which shift individual escape counts by at most two and alter no statistical conclusion.

The 50 seeds are subjected to four matched-compute frequency curricula:
\begin{description}
    \item[Standard Training (STD)] The control condition. All 120 epochs are trained directly on the target high-frequency landscape ($f=3.0$).
    \item[Forward Curriculum (FWD)] The spectral homotopy condition. The model is trained sequentially on ascending frequencies: Phase 1 ($f=1.0$), Phase 2 ($f=2.0$), and Phase 3 ($f=3.0$).
    \item[Reverse Curriculum (REV)] A descending sequence to test for irreversible high-frequency locking and trap irreversibility: Phase 1 ($f=3.0$), Phase 2 ($f=2.0$), and Phase 3 ($f=1.0$).
    \item[Random Curriculum (RND)] A noise control in which the three phase frequencies $\{1.0, 2.0, 3.0\}$ are presented in an independent, uniformly random order for each seed, to ensure that any observed benefits in the FWD condition derive from the directional spectral ordering, not merely exposure to diverse data distributions.
\end{description}
By continuously mapping the target accuracy and $F_x$ traces of these 50 seeds across the three phases, this design isolates the moment the latent scaffold engages, testing whether the curriculum transports the optimizer past the high-frequency Fourier locks.

\section{Results}\label{sec:results} 
Our experimental design isolated the optimization dynamics of the Data Re-uploading Parameterized Quantum Circuit (DRU-PQC) to determine why expressive architectures frequently fail to learn high-frequency targets. The results show that the primary bottleneck is not the representational capacity of the circuit, but the spectral traps created by FL during standard gradient descent.

\subsection{Diagnosing the Joint Bottleneck}\label{sec:bottleneck}
To confirm that the failure mode is driven by the non-linear coupling of encoding and entangling parameters (the joint bottleneck), we first analyzed the baseline behavior of the network under standard training and attempted to rescue it via routing ablations.

\textbf{The Mean-Prediction Trap:} Under standard training (STD), roughly 90\% of random initializations (88--94\% across independent experimental runs) collapsed into a degenerate mean-prediction trap, yielding test accuracies effectively at chance (near 0.50).

\begin{figure}[htbp]
    \centering
    \includegraphics[width=0.95\linewidth]{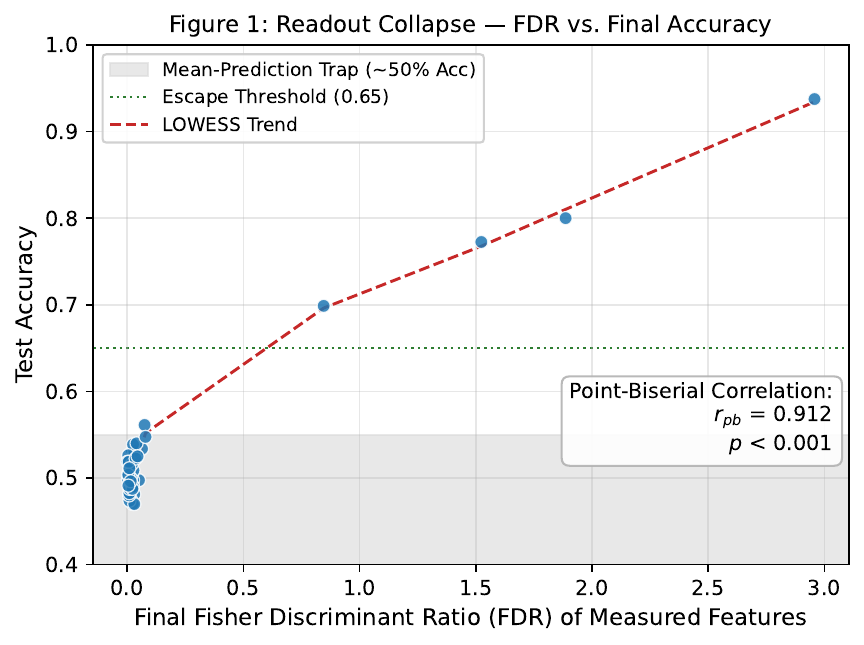}
    \caption{\textbf{Readout collapse as an optimization trap diagnostic.} Final Fisher discriminant ratio (FDR) of the measured features versus test accuracy for the 4-qubit DRU-PQC across the 50 seeds of the diagnostic-instrumented run (4/50 escapes), an independent seed set from the routing and curriculum experiments (Sec.~\ref{sec:expDesign}). The optimization landscape exhibits a bimodal distribution. Initializations that collapse into the degenerate mean-prediction trap (shaded region at $\sim$50\% accuracy) show FDR collapsing toward zero: the measured features become class-degenerate. Models that map the target geometry (surpassing the 0.65 escape threshold, green dotted line) maintain a non-zero FDR ($r_{pb} = 0.912$, $p < 0.001$). The parameter-space QFIM of the same circuits shows no corresponding collapse (Sec.~\ref{sec:bottleneck}), locating the failure at the readout's label alignment rather than in the state geometry. The dashed red line is a LOWESS trend.}
    \label{fig:qfi_diagnostic}
\end{figure}

\textbf{Readout Collapse and Intact State Geometry:} Because performance in this expressive landscape is bimodal---characterized by a dense cluster of trapped seeds at chance-level accuracy and a sparse distribution of successful escapes---continuous correlation metrics become statistically diluted by the random rank-ordering within the noise cluster. To quantify diagnostic power, we evaluated the point-biserial correlation ($r_{pb}$) between each candidate diagnostic and the binary escape outcome (final accuracy $> 0.65$). The FDR of the measured features discriminates trapped from escaped runs sharply ($r_{pb} = 0.912$, $p < 0.001$; Figure~\ref{fig:qfi_diagnostic}): trapped seeds exhibit FDR scores collapsing toward zero, the signature of a class-degenerate readout. The signal is also early, reaching $r_{pb} = 0.73$ by epoch 10 and $0.90$ by epoch 60, so the readout collapse is established long before training ends.

The parameter-space QFIM shows no such collapse. Across the same seeds, $\frac{1}{N}\text{Tr}(F)$ remains within a narrow band of order unity for trapped and escaped circuits alike ($0.86$--$1.18$ at the final epoch) and carries no diagnostic signal ($r_{pb} = 0.10$, $p = 0.51$). This is a substantive negative result: the gradient death in locked circuits is not caused by degeneracy of the parameter-space geometry, and diagnostics built on parameter-space QFIM rank---natural extrapolations of the static overparametrization theory~\cite{larocca2023theory} to training dynamics---do not detect the trap. What dies is the label alignment of the readout, not the responsiveness of the state.

The input-space QFI completes the dissociation. Its final value alone does not separate the populations ($r_{pb} = -0.19$, $p = 0.19$): trapped circuits oscillate strongly with the input throughout training. Its trajectory, however, does. Escaping circuits migrate their effective frequency content substantially over the run (mean $\Delta F_x = -30.5$, from $112$ to $82$ on the probe batch), while trapped circuits are spectrally frozen (mean $\Delta F_x = +1.0$); the correlation between $\Delta F_x$ and escape is $r_{pb} = -0.48$ ($p < 0.001$). A trapped circuit's spectral content simply does not move. We return to this signature, and its independent replication under the curriculum, in Sec.~\ref{sec:scaffold}.

\begin{figure}[htbp]
    \centering
    \includegraphics[width=0.9\linewidth]{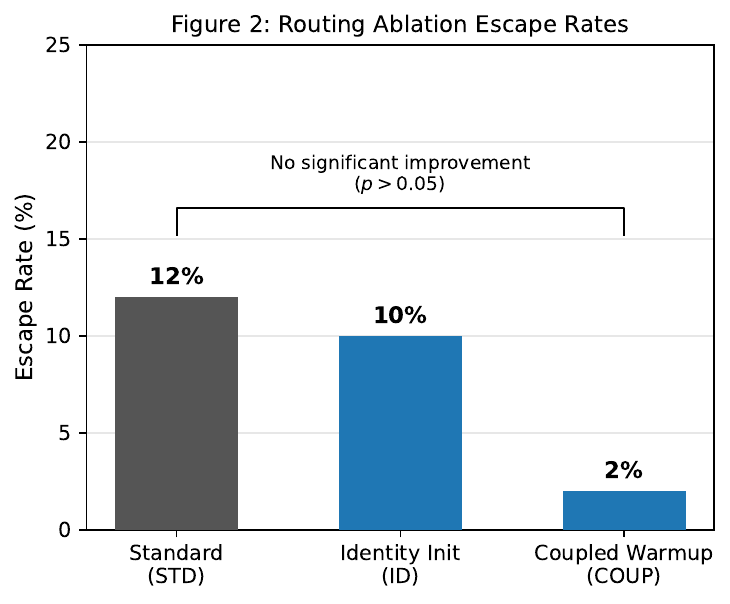}
\caption{\textbf{Routing ablation interventions fail to mitigate Fourier locking.} Comparison of optimizer escape rates from the mean-prediction trap across three distinct routing configurations (evaluated over 50 independent random initializations). Standard training (STD, 12\% escape rate) is compared against interventions explicitly designed to pre-align information routing: Identity Initialization (ID, 10\%) and Coupled Warmup (COUP, 2\%). Neither routing intervention yielded a statistically significant improvement over the baseline ($p > 0.05$, two-sided Fisher's exact tests); the coupled warmup in fact significantly reduced mean final accuracy (Welch's $t$-test, $p = 0.020$). This confirms that the joint parameter bottleneck is strictly dominated by the encoding initialization lottery; if the encoding weights ($\omega$) commit to a spurious high-frequency attractor, the entangling parameters ($\theta$) cannot route an uncaptured signal to rescue the model's predictive performance.}
\label{fig:routing_ablation}
\end{figure}

\textbf{Failure of Routing Ablation:} Initializing the entangling layers to Identity (ID) or applying an independent warmup optimization phase (COUP) failed to produce a statistically significant increase in the escape rate compared to the STD baseline (ID: 10\% vs.\ 12\%, $p = 1.00$; COUP: 2\% vs.\ 12\%, $p = 0.11$; two-sided Fisher's exact tests), as illustrated in Figure~\ref{fig:routing_ablation}. Notably, the coupled warmup not only failed to help but significantly reduced mean final accuracy relative to standard training (Welch's $t$-test, $p = 0.020$), suggesting that pre-aligning the entangling layers to a randomly initialized encoding state actively entrenches the misaligned attractor.

These results show that encoding weight initialization is the dominant variable in this type of training regime. The optimizer cannot rescue a detuned encoding layer by adjusting the entangling layers; if the encoding weights commit to the wrong Fourier frequency early in training, the signal effectively does not exist to route. That entangler-side pre-alignment can actively degrade performance further indicates the entangling parameters are downstream of, and captive to, the encoding configuration.

\subsection{Spectral Homotopy and Bimodal Probability Shifts}
Because performance in this expressive landscape is bimodal, we evaluate the intervention via the escape rate (final accuracy $> 0.65$) and analyze the pairwise comparisons using one-sided Fisher's exact tests. 

Standard (STD) training yielded a 6\% escape rate (3/50 seeds). The forward (FWD) sequence ($f=1.0 \rightarrow 2.0 \rightarrow 3.0$) increased this escape rate threefold to 18\% (9/50 seeds). While this represents a strong directional trend in mitigating the trap, it falls just short of conventional statistical significance ($p = 0.061$, one-sided Fisher's exact test, $\alpha = 0.05$). The random control (RND) yielded an 8\% escape rate (4/50 seeds), which is statistically indistinguishable from the standard baseline ($p = 0.50$, one-sided Fisher's exact test), indicating that unguided data diversity provides no routing benefit on average. Because each seed's phase order is drawn from a seeded generator, the per-seed RND orderings are exactly reconstructible, and stratifying by them provides an internal test of the ordering hypothesis. Twelve of the 50 RND seeds drew the exact forward order $[1.0, 2.0, 3.0]$ by chance, and three of RND's four escapes occurred among them (25\% vs.\ 2.6\% for the remaining orderings; $p = 0.038$, one-sided Fisher's exact test); the fourth escape drew $[2.0, 1.0, 3.0]$, which also approaches the target from below. No seed whose sequence did not end at the target frequency escaped (0/33; $p = 0.010$). Correctly stratified, the noise control therefore corroborates the ordering effect rather than merely failing to show one.

\subsubsection{The Reverse Curriculum: Attractor Collapse and Spectral Forgetting} 
\begin{figure}[htbp]
    \centering
    \includegraphics[width=0.9\linewidth]{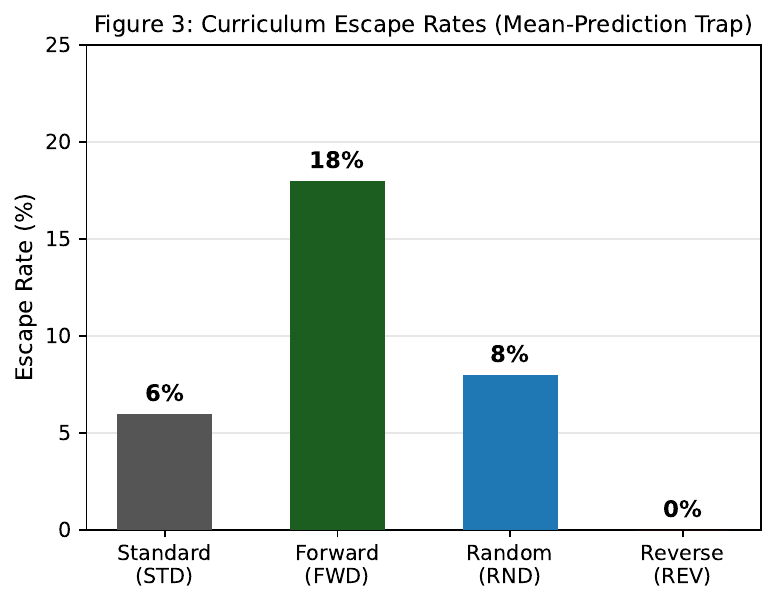}
    \caption{\textbf{Impact of spectral curricula on Fourier lock escape rates.} Successful initialization rates (final accuracy $> 0.65$) across 50 independent seeds for the 4-qubit DRU-PQC under the four matched-compute curricula. Standard training (STD) establishes a 6\% baseline escape rate in this seed set. The forward spectral homotopy (FWD) triples the escape rate to 18\% by progressively convexifying the loss landscape during early optimization. Unguided data diversity (labeled as RND) yields no statistical improvement on average (8\%; see text for the stratified analysis of its internally drawn orderings), and the reverse curriculum (REV) produces a 0\% escape rate, consistent with immediate high-frequency attractor collapse followed by spectral forgetting. Routing-side interventions were separately shown to provide no rescue (Fig.~\ref{fig:routing_ablation}). Together these results support the interpretation that Fourier locking is a spectral trap in the joint parameter space, and that frequency staging is necessary to circumvent it.}
    \label{fig:escape_rates}
\end{figure}

The Reverse (REV) sequence ($f=3.0 \rightarrow 2.0 \rightarrow 1.0$) tests the irreversibility of the high-frequency trap. Evaluating the internal phase logs of these 50 seeds reveals a dual failure mode that guarantees a 0\% final escape rate, shown in Figure~\ref{fig:escape_rates}.

For the vast majority of initializations (94\%, 47/50 seeds), exposure to the $f=3.0$ landscape in Phase 1 triggered immediate, irreversible attractor collapse. The encoding weights fell into degenerate minima, and accuracy remained at chance levels for the entire 120-epoch duration.

However, the remaining 6\% (3/50 seeds) fit the high-frequency boundary directly in Phase 1 (achieving target accuracies of $0.850$, $0.686$, and $0.901$). Yet, when the curriculum subsequently forced the optimizer into the lower-frequency geometries of Phases 2 and 3, this knowledge was overwritten. By the end of Phase 3, the accuracy of these previously successful seeds had completely collapsed back to chance levels (e.g., descending $0.901 \rightarrow 0.514 \rightarrow 0.504$). We term this secondary failure mode catastrophic spectral forgetting. It demonstrates that the DRU-PQC architecture lacks the capacity to retain competing frequency representations; even if a model successfully bypasses the initial attractor collapse, exposing the parameters to misaligned, descending frequencies destroyed the previously secured narrow-band attractor in all three observed cases.

\subsection{Distinguishing Spectral Homotopy from Classical Curriculum Transfer}
Given the harmonic relationship of the target datasets ($f=1.0 \rightarrow 2.0 \rightarrow 3.0$), a natural concern is that the observed Phase 3 spike may simply reflect classical curriculum transfer—where a model reuses previously learned geometric features—rather than a spectral regularization of the quantum parameter space. However, the mathematical structure of the DRU-PQC rules out a purely classical transfer interpretation.

Unlike classical neural networks, which possess flexible feature extractors, the DRU-PQC constructs a truncated Fourier series (as defined in Equation \ref{eq:fourier}). The accessible frequency spectrum of the circuit is bounded by the physical state of the encoding weights ($\omega$). Therefore, for the circuit to successfully fit the $f=3.0$ boundary in Phase 3, the encoding weights must physically realign to the $f=3.0$ parameter basin; the entangling layers ($\theta$) cannot simply "reuse" a low-frequency representation to map a high-frequency boundary.

The bimodal experimental data clarifies this mechanical distinction. If the optimizer could reliably find the $f=3.0$ encoding basin through standard gradient descent, the STD condition would not exhibit a 94\% failure rate. The failure of STD training proves that the $f=3.0$ landscape is too non-convex and oscillatory, causing the encoding weights to irreversibly collapse into spurious local minima.

The success of the forward (FWD) sequence demonstrates that the harmonic progression acts as a true spectral homotopy. The smooth $f=1.0$ landscape pulls the joint parameters $(\omega, \theta)$ into a broad, generalized basin. As the target frequency scales to $f=2.0$ and finally $f=3.0$, the loss basin physically tightens and becomes oscillatory. By scaffolding the parameters in the early phases, the encoding weights are positioned within the correct spectral funnel, allowing them to smoothly track the shrinking basin and realign to the $f=3.0$ target. The resulting Phase-3 growth in $F_x$ is therefore not a measure of classical feature transfer, but the geometric signature of the encoding weights realigning into the high-frequency Fourier basin that was inaccessible from random initialization.

\subsection{The Latent Scaffold: Trapped Circuits Are Spectrally Frozen, Escapees Migrate}
\label{sec:scaffold}
We analyzed the target accuracy ($f=3.0$) and input-QFI trajectories of the FWD seeds across the three training phases to identify how the forward curriculum constructs a viable encoding alignment. If the curriculum functioned via classical feature extraction or harmonic overlap, we would expect steady, incremental accuracy gains across the low-frequency phases.

\begin{figure}[htbp]
    \centering
    \includegraphics[width=0.95\linewidth]{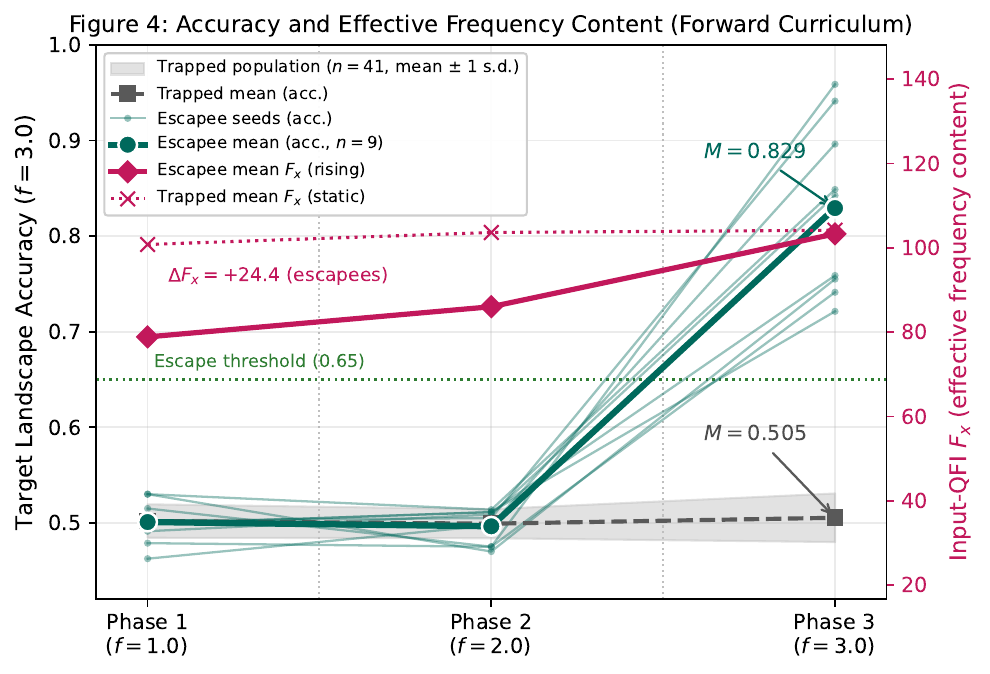}
    \caption{\textbf{The latent scaffold in accuracy and effective frequency content.} Target accuracy ($f=3.0$, left axis) and input-QFI $F_x$ (right axis) over the 120-epoch forward curriculum (FWD), in three 40-epoch spectral phases ($f=1.0 \rightarrow 2.0 \rightarrow 3.0$) divided by vertical dashed lines. Accuracy of the successful initializations follows a step function: chance through Phases 1--2 ($M \approx 0.50$), indistinguishable in accuracy from trapped runs, then a jump to $M = 0.829$ at Phase 3. The $F_x$ trajectories separate the populations where accuracy cannot: escapees grow $F_x$ monotonically in step with the curriculum (8 of 9 rising; $d = 1.34$, $p < 0.001$), while trapped seeds hold a static effective frequency throughout. Both populations converge to the same $F_x$ magnitude by Phase 3; the distinguishing variable is the trajectory, not the endpoint. The low-frequency phases do not teach the decision boundary---they transport the encoding into the target frequency band.}
    \label{fig:latent_scaffold}
\end{figure}

The accuracy data, visualized in Figure~\ref{fig:latent_scaffold}, show a near-perfect step function instead. Across the successful FWD seeds, mean target accuracy remained at chance through the low-frequency phases---Phase 1 ($M = 0.501$), Phase 2 ($M = 0.496$)---then jumped to $M = 0.829$ at Phase 3 onset, while trapped seeds stayed at chance ($M = 0.505$). Individual trajectories (e.g., $0.502 \rightarrow 0.505 \rightarrow 0.941$) confirm there is no feature-transfer drift during Phase 2, and the flat pre-Phase-3 accuracy of eventual escapees is indistinguishable from the trapped population, ruling out survivorship bias in the accuracy signal.

The $F_x$ trajectories, by contrast, are not flat.

\textbf{Escapees climb the curriculum's frequency ladder; trapped circuits do not move.} Under FWD, the escapees' $F_x$ grows monotonically across the phases ($\Delta F_x = +24.4 \pm 25.6$ from Phase 1 to Phase 3; 8 of 9 escapees rising), while the trapped population is spectrally static ($\Delta F_x = +3.4 \pm 12.9$); the separation is the largest effect in our data ($d = 1.34$, $r_{pb} = 0.46$, $p < 0.001$). The same dichotomy appears, out of sample, in the independent direct-training experiment of Sec.~\ref{sec:bottleneck}: there, escaping circuits migrate their frequency content downward toward the target band ($\Delta F_x = -30.5$ on average) while trapped circuits are again frozen ($r_{pb} = -0.48$, $p < 0.001$). The direction of migration differs with the training regime, but the signature---spectral motion versus spectral immobility---replicates across independent seed sets. Notably, FWD escapees and trapped seeds converge to the same $F_x$ magnitude by Phase 3 ($103.3$ vs.\ $104.1$): the distinguishing variable is not the final frequency content but its history. Trapped circuits hold a misaligned effective frequency from initialization onward; escaping circuits are transported, phase by phase, into alignment with the target band. This is the behavior the latent-scaffold hypothesis predicts: the low-frequency phases do not teach the decision boundary---accuracy stays at chance---but they hold the encoding in a trainable configuration and raise its frequency content in step with the target, so that when Phase 3 begins the encoding weights snap into the high-frequency basin that was inaccessible from random initialization.

\textbf{Initial frequency content as a lottery variable.} The initialization lottery itself also leaves a spectral trace, though a less robust one. A seed's early-training $F_x$ is a stable property of its initialization within a fixed condition: across the four curriculum conditions, which train on different frequencies in Phase 1, per-seed Phase-1 $F_x$ values correlate at Spearman $\rho = 0.86$--$0.91$. Pooling at the seed level (per-seed mean of Phase-1 $F_x$ across the four conditions), the 14 initializations that escape under at least one curriculum carry lower initial frequency content than the 36 that never escape ($F_x = 79.4$ vs.\ $103.7$; $d = -0.83$, $r_{pb} = -0.35$, $p = 0.012$, $n = 50$ shared initializations); the association transfers across conditions ($|r_{pb}|$ up to $0.31$ between one condition's Phase-1 $F_x$ and another condition's escape outcome); and the three initializations that fit the $f=3.0$ boundary directly in the reverse curriculum's first phase are precisely low-$F_x$ outliers ($F_x = 62.7$ vs.\ $101.3$ for the remainder; one-sided Mann--Whitney $p = 0.010$). The out-of-sample check on this effect is mixed, however: in the independent diagnostic experiment, the epoch-0 association replicates under the identity-initialized condition ($r_{pb} = -0.30$, $p = 0.03$, with the correlation stable in magnitude at every tracked epoch) but not under standard initialization ($r_{pb} = +0.07$, $p = 0.62$, with only four escape events). The condition dependence has a visible cause: although the encoding weights are drawn identically in the two conditions, their epoch-0 $F_x$ values are uncorrelated across seeds (Spearman $\rho = 0.02$), because the entangler configuration reshapes the effective frequency content of the multi-qubit state. We therefore report the initial-$F_x$ association as suggestive rather than established; the replicated signature of Fourier locking is the spectral immobility of the trapped circuits.

We note that the lottery statistics derive from the 50 shared initializations of the curriculum experiment and a directional hypothesis formed after inspecting the FWD trajectories; the within-condition Phase-1 correlations are individually weak and uneven ($r_{pb} = -0.17$ under STD, $-0.27$ under FWD, $-0.04$ under RND; undefined under REV, which has no escapes), and it is the pooled seed-level test and the cross-condition transfer that carry the claim.

\section{Conclusion and Future Work}
\subsection{Conclusion}

As discussed in Section~\ref{sec:background}, Schuld et al.~\cite{schuld2021effect} establish that the accessible frequency spectrum of a DRU-PQC is determined by the encoding Hamiltonian eigenvalues, and that frequency mismatch causes training failure as an expressivity limitation. This study reframes the optimization bottleneck in expressive DRU-PQCs: by measuring the Fisher geometry of the trained circuits directly, we show that models fail not through chaotic landscape flattening (standard barren plateaus), and not through a collapse of the parameter-space information geometry, but through FL---a spectral misalignment in which a fully responsive state oscillates with the input at frequencies the readout cannot map to the class labels.

An algebraic theory of QFIM rank already exists for periodic-structured QNNs: Larocca et al.~\cite{larocca2023theory} relate the achievable QFIM rank to the dimension of the dynamical Lie algebra and define overparametrization as the saturation of that rank. That framework characterizes a static, architectural property of the ansatz, and a natural dynamic extrapolation would predict that trained-in failure appears as rank collapse. Our measurements rule this out for the re-uploading regime: trapped circuits retain a non-degenerate parameter-space QFIM throughout training. The diagnostic content lives elsewhere---in the input-space QFI $F_x$, which reads out the effective frequency of the encoding, and in the Fisher discriminant ratio of the measured features, which reads out its alignment with the labels. Measured through $F_x$, the locking is literal: in both independent experiments, trapped circuits hold a frozen, misaligned frequency content from initialization onward, while every escaping population migrates its spectral content---downward under direct training, upward in step with the curriculum. Whether the initial value of $F_x$ predicts escape before training begins appears condition-dependent in our data, and we report it as a suggestive effect.

The application of frequency-staged homotopy provided empirical support for this mechanism during training: by convexifying the early loss landscape, we delayed the onset of high-frequency attractors, and the escaping circuits grew their effective frequency content monotonically in step with the curriculum---the largest effect we measure---which tripled the optimizer's escape rate.

Nevertheless, the practical limitations of this intervention are instructive. Even with optimal frequency pacing, 82\% of random initializations remained permanently trapped in the degenerate mean-prediction state. This fragility underscores a central characteristic of expressive QML landscapes: they cannot be consistently navigated by gradient descent from random starting points. While frequency pacing strongly suggests the existence and mechanics of a latent scaffold, the 82\% failure rate confirms that the encoding initialization \textit{lottery} is too dominant to be solved by optimization curricula alone.

Therefore, future research requires new methods other than random initialization. For instance, strategies utilizing classical spectral analysis to deterministically pre-align quantum encoding weights, guided by the Fisher diagnostics introduced here, represent the most obvious next step toward making expressive DRU-PQCs reliably trainable for high-frequency target landscapes.

\subsection{Future Work}
While spectral homotopy appears to mitigate FL, the bimodal distribution of outcomes reveals a fundamental limitation: 82\% of random initializations still failed to escape the trap under the forward curriculum. This indicates that the sensitivity to encoding initialization is mitigated, but not entirely resolved, by frequency pacing alone. Below we have enumerated several strategies for improving quantum classifier performance using DRU-PQCs. 

\begin{itemize}
    \item \textbf{Deterministic Encoding Initialization:} Because entangling optimization cannot rescue a detuned encoding layer, future work should investigate alternatives to random initialization. Utilizing classical spectral analysis (e.g. applying a fast Fourier transform to a classical dataset prior to encoding) would allow us to deterministically pre-set the quantum encoding weights $\omega$. This could ensure the circuit begins within a viable spectral basin, substantially reducing initialization dependency. On the other hand, if such pre-computation was inaccurate, then it could lead to more variance in the FL escape rates. 
    \item \textbf{$F_x$-Guided Homotopy:} In this study, spectral homotopy was applied in fixed 40-epoch phases. Because the input-QFI $F_x$ tracks the engagement of the latent scaffold directly, future algorithms could utilize $F_x$ as an active phase trigger. The curriculum would transition to higher frequencies ($L_{f_i} \rightarrow L_{f_{i+1}}$) only once the effective frequency content has migrated into the current band, ensuring the encoding is sufficiently scaffolded before landscape complexity increases.
    \item \textbf{Scaling and Hardware Noise:} This analysis was conducted under noise-free simulations. Extending this framework to wider multi-qubit systems and deeper re-uploading architectures on noisy intermediate-scale quantum (NISQ) hardware is relevant to this investigation since hardware noise inherently dampens high-frequency Fourier components. Investigating how device noise interacts with or potentially regularizes FL will be critical for bridging this theoretical framework to physical implementations now and in the fault-tolerant era.
\end{itemize}

\begin{acknowledgments}
The author thanks James Wootton for his continued guidance and Daniel Bultrini for his feedback.
\end{acknowledgments}

\section*{Competing Interests}
The author is Chief Technology Officer of Moth. The author declares no other competing interests.

\section*{Data Availability}
The simulation code, experiment data, and figure-generation notebooks supporting the findings of this study are openly available at \url{https://github.com/moth-quantum/fourier_locking_study}.
\appendix

\section{Geometric Distinguishability and QFIM Derivation}
\label{app:qfi}

The QFIM and its interpretation as a measure of state-space distinguishability are standard~\cite{larocca2023theory}. We recall the Bures-distance derivation, which follows established constructions~\cite{meyer2021fisher,liu2020quantum}, to make the present argument self-contained and to extend it past the rank interpretation to a limiting-case theorem: QFIM collapse, were it to occur, would force gradient death. Our measurements show this regime is not realized in Fourier-locked circuits (Sec.~\ref{sec:results}); we include the derivation because it sharpens the negative result. The theorem tells us what parameter-space severing would look like, and the data tell us the locked circuits look nothing like it. The Bures metric measures the statistical distinguishability between two infinitesimally close quantum states, allowing us to isolate genuine geometric movement from physically unobservable parameter updates.

\subsection{Fidelity and the Taylor Expansion}
Consider a pure quantum state $|\psi(\theta)\rangle$ parameterized by $\theta$. If the optimizer updates the parameters by an infinitesimal amount $d\theta$, the distinguishability between the original state and the updated state $|\psi(\theta + d\theta)\rangle$ is measured by their squared overlap, or Fidelity ($\mathcal{F}$):
\begin{equation}
\mathcal{F} = |\langle \psi(\theta) | \psi(\theta + d\theta) \rangle|^2
\end{equation}
Next we expand the perturbed state to the second order in $d\theta$:
\begin{equation}
|\psi(\theta + d\theta)\rangle \approx |\psi\rangle + \sum_i |\partial_i \psi\rangle d\theta_i + \frac{1}{2}\sum_{i,j} |\partial_i \partial_j \psi\rangle d\theta_i d\theta_j
\end{equation}
Defining the first-order differential as $|d\psi\rangle \equiv \sum_i |\partial_i \psi\rangle d\theta_i$ and the unscaled second-order differential as $|d^2\psi\rangle \equiv \sum_{i,j} |\partial_i \partial_j \psi\rangle d\theta_i d\theta_j$, this expansion can be written as:
\begin{equation} \label{eq:perturbed}
|\psi(\theta + d\theta)\rangle \approx |\psi\rangle + |d\psi\rangle + \frac{1}{2}|d^2\psi\rangle
\end{equation}

\subsection{State Normalization Constraints}
Because quantum states must represent normalized probabilities ($\langle \psi | \psi \rangle = 1$), the expansion of the perturbed state is heavily constrained. Evaluating the normalization of the perturbed state to the second order yields: 

\begin{equation}
1 \approx \langle \psi + d\psi + \frac{1}{2}d^2\psi | \psi + d\psi + \frac{1}{2}d^2\psi \rangle
\end{equation}

\begin{equation}
\begin{split}
1 \approx 1 &+ \left( \langle \psi | d\psi \rangle + \langle d\psi | \psi \rangle \right) + \langle d\psi | d\psi \rangle \\
  &+ \frac{1}{2} \left( \langle \psi | d^2\psi \rangle + \langle d^2\psi | \psi \rangle \right)
\end{split}
\end{equation}

By isolating the terms by their differential order and recognizing that the unperturbed state is already normalized (yielding the leading 1), this expansion imposes two fundamental constraints on the state derivatives:

\begin{itemize}
    \item \textbf{First-order:} $\langle \psi | d\psi \rangle + \langle d\psi | \psi \rangle = 0 \implies 2\,\text{Re}\langle \psi | d\psi \rangle = 0$. This dictates that the real part of the first-order overlap must be zero. (Note that the imaginary part remains unconstrained, a freedom that manifests physically as the Berry phase).
    \item \textbf{Second-order:} $\frac{1}{2} \left( \langle \psi | d^2\psi \rangle + \langle d^2\psi | \psi \rangle \right) = - \langle d\psi | d\psi \rangle$. This dictates that the real part of the second-order overlap equals the negative of the squared norm of the first-order derivative state.
\end{itemize}
\subsection{Bures Distance Formulation}

The Bures distance between two close pure states is
\begin{equation}
d_B^2 = 2(1 - \sqrt{\mathcal{F}}),
\end{equation}
where $\mathcal{F} = |\langle \psi(\theta) | \psi(\theta + d\theta) \rangle|^2$ is the fidelity. Expanding the fidelity using the perturbed state \eqref{eq:perturbed} yields
\begin{multline}
\mathcal{F} = \left( 1 + \langle \psi | d\psi \rangle + \tfrac{1}{2} \langle \psi | d^2\psi \rangle \right) \\
              \times \left( 1 + \langle d\psi | \psi \rangle + \tfrac{1}{2} \langle d^2\psi | \psi \rangle \right)
              + \mathcal{O}(d\theta^3).
\end{multline}
Multiplying out and retaining terms to second order in $d\theta$:
\begin{multline}
\mathcal{F} \approx 1 + \left( \langle \psi | d\psi \rangle + \langle d\psi | \psi \rangle \right)
                  + |\langle \psi | d\psi \rangle|^2 \\
                  + \tfrac{1}{2}\left( \langle \psi | d^2\psi \rangle + \langle d^2\psi | \psi \rangle \right).
\end{multline}
Applying the first-order and second-order normalization constraints derived above, the first-order term vanishes and the second-order term equals $-\langle d\psi | d\psi \rangle$, giving
\begin{equation}
\mathcal{F} \approx 1 - \langle d\psi | d\psi \rangle + |\langle \psi | d\psi \rangle|^2.
\end{equation}
Since $\mathcal{F}$ is close to unity, we Taylor-expand the square root as $\sqrt{\mathcal{F}} \approx 1 - \tfrac{1}{2}(1 - \mathcal{F})$, so that
\begin{equation}
d_B^2 \approx 2\left(1 - \sqrt{\mathcal{F}}\right) \approx 1 - \mathcal{F}.
\end{equation}
Substituting the second-order expression for $\mathcal{F}$ gives the infinitesimal Bures distance
\begin{equation}
ds^2 \equiv d_B^2 \approx \langle d\psi | d\psi \rangle - |\langle \psi | d\psi \rangle|^2,
\label{eq:ds2}
\end{equation}
where the first term captures the movement of the state vector in Hilbert space and the second subtracts the component parallel to $|\psi\rangle$—the physically unobservable $U(1)$ gauge direction. Equation~(\ref{eq:ds2}) is the geometric object from which the QFIM elements are extracted in the next step.

\subsection{Extraction of the QFIM}
In quantum estimation theory, the distance between two infinitesimally close pure states is related to the QFIM $F$ by a conventional scaling factor of 4~\cite{meyer2021fisher,liu2020quantum}. Substituting the explicit derivative sum $|d\psi\rangle = \sum_i |\partial_i \psi\rangle d\theta_i$ into our derived distance equation yields:
\begin{equation}
ds^2 = \sum_{i,j} \left( \langle \partial_i \psi | \partial_j \psi \rangle - \langle \psi | \partial_i \psi \rangle \langle \partial_j \psi | \psi \rangle \right) d\theta_i d\theta_j
\end{equation}
To serve as a valid metric tensor, the QFIM must be a real, positive semi-definite matrix. We therefore take the real part of this distance (the imaginary component corresponds to the geometrically distinct Berry curvature~\cite{berry1984quantal,zanardi2007information}). Because $\overline{\langle a|b\rangle} = \langle b|a\rangle$, the second term $\langle \psi | \partial_i \psi \rangle \langle \partial_j \psi | \psi \rangle$ is the exact complex conjugate of $\langle \partial_i \psi | \psi \rangle \langle \psi | \partial_j \psi \rangle$. Under the real projection $\text{Re}(\cdots)$ these terms are equivalent. 

Applying the scaling factor of 4 yields the standard form of the QFIM elements utilized in this study:
\begin{equation}
F_{ij} = 4\,\text{Re}\!\left( \langle \partial_i \psi | \partial_j \psi \rangle - \langle \partial_i \psi | \psi \rangle \langle \psi | \partial_j \psi \rangle \right)
\end{equation}

\subsection{Gradient Vanishing from QFIM Collapse} \label{Sec:GradientVanishing}
A QFIM collapse ($F_{ii} \to 0$) mathematically guarantees the death of the analytical gradient ($\nabla_{\theta_i} L \to 0$) for any loss function derived from a physical observable $H$.

\textbf{Step 1:} The gradient of the loss $L = \langle\psi|H|\psi\rangle$ with respect to parameter $\theta_i$ is given by $\partial_i L = 2\,\text{Re}\langle \partial_i\psi | H | \psi \rangle$. To establish a bound, we introduce a centering argument by subtracting the expectation value $\langle H \rangle$:
\begin{equation}
\langle\partial_i\psi|H|\psi\rangle = \langle\partial_i\psi|(H - \langle H\rangle)|\psi\rangle + \langle H\rangle\langle\partial_i\psi|\psi\rangle
\end{equation}
Because $H$ is a physical observable (Hermitian), $\langle H \rangle$ is strictly real. Applying the normalization constraint ($\text{Re}\langle\partial_i\psi|\psi\rangle = 0$), the second term $\langle H\rangle\langle\partial_i\psi|\psi\rangle$ is entirely imaginary. Because the gradient takes only the real part, this uncentered term vanishes completely, leaving:
\begin{equation}
\partial_i L = 2\,\text{Re}\langle\partial_i\psi|(H - \langle H\rangle)|\psi\rangle
\end{equation}

\textbf{Step 2:} Applying the Cauchy-Schwarz inequality to this centered Hilbert space inner product provides a strict upper bound via the variance of the observable:
\begin{equation}
|\partial_i L|^2 = 4\,|\text{Re}\langle\partial_i\psi|(H-\langle H\rangle)|\psi\rangle|^2 \leq 4\,\langle\partial_i\psi|\partial_i\psi\rangle \cdot \text{Var}(H)
\end{equation}
Notice that this bound relies on the uncentered norm $\langle\partial_i\psi|\partial_i\psi\rangle$.

\textbf{Step 3:} The diagonal elements of the QFIM are given by $F_{ii} = 4(\langle\partial_i\psi|\partial_i\psi\rangle - |\langle\psi|\partial_i\psi\rangle|^2)$. Since $\langle\partial_i\psi|\partial_i\psi\rangle \geq |\langle\psi|\partial_i\psi\rangle|^2$ by the Cauchy-Schwarz inequality (with $\langle\psi|\psi\rangle = 1$), the quantity $F_{ii}$ is non-negative.

When $F_{ii} = 0$, it enforces the equality $\langle\partial_i\psi|\partial_i\psi\rangle = |\langle\psi|\partial_i\psi\rangle|^2$. If $\langle\partial_i\psi|\partial_i\psi\rangle = 0$, the state derivative is exactly zero and the gradient vanishes trivially. Otherwise, by the equality condition of the Cauchy-Schwarz inequality ($|\langle u|v \rangle|^2 = \langle u|u \rangle\langle v|v \rangle \iff |u\rangle \propto |v\rangle$), this strict equality forces the derivative state to be entirely parallel to the current state. Applying the normalization constraint $\text{Re}\langle\psi|\partial_i\psi\rangle = 0$, the constant of proportionality must be purely imaginary:
\begin{equation}
|\partial_i\psi\rangle = i\phi\,|\psi\rangle \quad \text{for some } \phi \in \mathbb{R}
\end{equation}

\textbf{Step 4:} Taking the Hermitian conjugate yields the corresponding bra: $\langle\partial_i\psi| = -i\phi\langle\psi|$. While the Cauchy-Schwarz bound in Step 2 remains loose (since $\langle\partial_i\psi|\partial_i\psi\rangle = \phi^2 \neq 0$), substituting this purely imaginary phase vector directly back into the exact uncentered gradient expression evaluates it identically to zero:
\begin{equation}
\partial_i L = 2\,\text{Re}(-i\phi\langle\psi|H|\psi\rangle) = -2\phi\langle H\rangle\,\text{Re}(i) = 0
\end{equation}

\textbf{Step 5:} If the full QFIM collapses to rank zero, then every diagonal element $F_{ii} = 0$. For any positive semi-definite matrix, the off-diagonal elements are bounded by the determinant inequality $|F_{ij}|^2 \leq F_{ii}F_{jj}$. 

Consequently, zero diagonal elements mathematically guarantee that the entire matrix is identically zero ($F=0$). This applies the argument across all parameter dimensions, guaranteeing a global vanishing gradient $\nabla_\theta L = 0$.

The derivation establishes an exact implication: if the QFIM collapses, the analytical gradient must vanish, with the parameter updates absorbed entirely into the unobservable $U(1)$ gauge direction. The converse does not hold, and our measurements show that Fourier-locked circuits realize the converse case. Their gradients die while $\frac{1}{N}\text{Tr}(F)$ remains of order unity: the state moves freely and observably under parameter variation, but the loss---which depends on the class alignment of the measured features rather than on state motion per se---has flattened for spectral reasons the parameter-space geometry cannot see. The gradient death of the Fourier lock is therefore not an instance of this theorem. We retain the theorem because it closes off the natural alternative explanation: a reader of the geometry might reasonably suspect gauge absorption, and the data, read against this derivation, exclude it.

\section{Quantum Circuit Architecture and Training Hyperparameters}
\label{app:arch}
\subsection{Gate-Level Architecture}
The 4-qubit Data Re-uploading Parameterized Quantum Circuit (DRU-PQC) acts on the initial state $|0\rangle^{\otimes 4}$ through a sequence of $L=2$ layers. Each layer $l$ applies a unitary transformation defined by a data-encoding operation and a trainable strongly entangling block.

\begin{figure*}[htbp]
    \centering
    \includegraphics[width=0.9\linewidth]{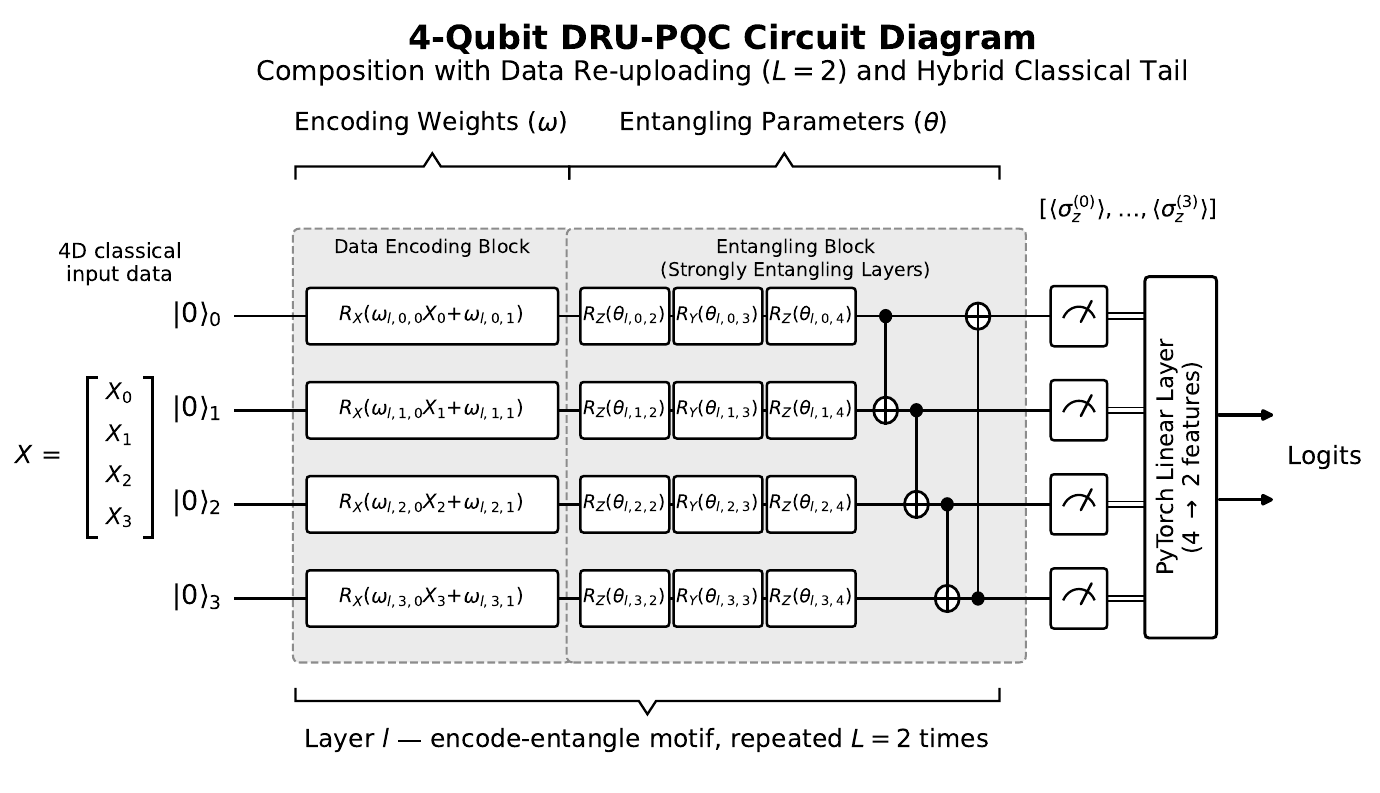}
    \caption{\textbf{Architecture of the 4-qubit DRU-PQC.} The classical 4D 
    input $X$ is encoded into the quantum state via Pauli-X rotations ($R_X$) 
    parameterized by trainable encoding weights ($\omega$), which act as frequency 
    selectors. An entangling block of $R_Z R_Y R_Z$ rotations ($\theta$) and a 
    CNOT ring follows, functioning as signal routers. This encode-entangle motif 
    repeats across $L=2$ layers, producing the nonlinear coupling between $\omega$ 
    and $\theta$ that underlies Fourier locking. The final state is read out via 
    local Pauli-Z expectation values ($\langle \sigma_z \rangle$) on all qubits 
    and passed through a classical linear layer to produce binary classification 
    logits.}
    \label{fig:dru_circuit}
\end{figure*}

\textbf{Encoding Block:} The classical 4D input $X$ is scaled by the trainable encoding weights and loaded into the circuit via $R_X$ Pauli rotations. For each qubit $i$ in layer $l$, the encoding operation combines a frequency-scaling weight ($\omega_{l, i, 0}$) and a phase bias ($\omega_{l, i, 1}$):
\begin{equation}
R_X(\omega_{l, i, 0} \cdot X_i + \omega_{l, i, 1})
\end{equation}


\textbf{Entangling Block:} Following the encoding, a Strongly Entangling Layers template is applied. This consists of parameterized local rotations $R_Z R_Y R_Z$ utilizing three per-qubit rotation angles $\theta_{l,i,k}$, occupying slots $k \in \{2, 3, 4\}$ of the five-parameter per-qubit weight vector, whose slots $0$ and $1$ hold the encoding weights $(\omega, \omega')$, followed by a hardware-efficient entanglement ring of CNOT gates.

\subsection{Measurement and Hybrid Layer}
The quantum state is evaluated by measuring the expectation value of the Pauli-Z observable on all 4 qubits:
\begin{equation}
[\langle \sigma_z^{(0)} \rangle, \langle \sigma_z^{(1)} \rangle, \langle \sigma_z^{(2)} \rangle, \langle \sigma_z^{(3)} \rangle]
\end{equation}
These four continuous values are then passed through a classical PyTorch linear layer ($4 \rightarrow 2$ features) to generate the final logits for binary classification, as shown in Figure~\ref{fig:dru_circuit}.

\subsection{Hyperparameter Configuration}
To ensure the bimodal probability space and Fisher diagnostic traces reported in Section~\ref{sec:results} are completely reproducible, the exact computational parameters utilized across all 50 seeds are detailed in Table \ref{tab:hyperparams}.

\begin{table}[htbp]
\caption{\label{tab:hyperparams} Experimental Hyperparameters}
\begin{ruledtabular}
\begin{tabular}{lc}
\textbf{Parameter} & \textbf{Value}\\
\hline
Qubit Count & 4 \\
Re-uploading Layers ($L$) & 2 \\
Trainable Quantum Parameters & 40 (shape: $2 \times 4 \times 5$) \\
Trainable Classical Parameters & 10 (Linear(4, 2)) \\
Total Trainable Parameters & 50 \\
Total Samples & 4000 \\
Batch Size & 256 \\
Optimizer & Adam (PyTorch backend) \\
Learning Rate ($\eta$) & 0.01 \\
Total Epochs & 120 (40 per curriculum phase) \\
Escape Threshold & Accuracy $> 0.65$ \\
Parameter-QFIM Scalar & $\frac{1}{N} \text{Tr}(F)$, fixed 8-input probe \\
Input-QFI ($F_x$) & Eq.~(\ref{eq:xqfi}), fixed 8-input probe \\
\end{tabular}
\end{ruledtabular}
\end{table}

\balance
\bibliography{references}

\end{document}